\documentclass[%
reprint,
showpacs,
preprintnumbers,
 amsmath,amssymb,
aps,
prl,
superscriptaddress
]{revtex4-1}

\usepackage{graphicx}
\usepackage{dcolumn}
\usepackage{bm}

\newcommand{\ofx}{(\bm{x},t)}

\newcommand{\bs}{\boldsymbol}

\usepackage[dvipsnames]{xcolor}

\definecolor{LinkBlue}{RGB}{46,46,145}
\usepackage{hyperref}
\hypersetup{
	colorlinks=true,
	linkcolor={LinkBlue},
	citecolor={LinkBlue},
	urlcolor={LinkBlue}
}

\newcounter{myequation}
\makeatletter
\@addtoreset{equation}{myequation}
\makeatother
\newcounter{myfigure}
\makeatletter
\@addtoreset{figure}{myfigure}
\makeatother

\begin{document}

\title{Large-Scale Pattern Formation in the Presence of Small-Scale Random Advection}

\author{Gregor Ibbeken}
\affiliation{
	Max Planck Institute for Dynamics and Self-Organization (MPI DS),
	Am Fa\ss berg 17, 37077 G\"ottingen, Germany
}
\affiliation{Faculty of Physics, University of G\"{o}ttingen, Friedrich-Hund-Platz 1, 37077 G\"{o}ttingen, Germany
}
\author{Gerrit Green}
\affiliation{
	Max Planck Institute for Dynamics and Self-Organization (MPI DS),
	Am Fa\ss berg 17, 37077 G\"ottingen, Germany
}
\affiliation{Faculty of Physics, University of G\"{o}ttingen, Friedrich-Hund-Platz 1, 37077 G\"{o}ttingen, Germany
}
\author{Michael Wilczek}
\email{michael.wilczek@ds.mpg.de}
\affiliation{
Max Planck Institute for Dynamics and Self-Organization (MPI DS),
Am Fa\ss berg 17, 37077 G\"ottingen, Germany
 }
\affiliation{Faculty of Physics, University of G\"{o}ttingen, Friedrich-Hund-Platz 1, 37077 G\"{o}ttingen, Germany
}

\begin{abstract}

Despite the presence of strong fluctuations, many turbulent systems such as Rayleigh-B\'enard convection and Taylor-Couette flow display self-organized large-scale flow patterns. How do small-scale turbulent fluctuations impact the emergence and stability of such large-scale flow patterns? Here, we approach this question conceptually by investigating a class of pattern forming systems in the presence of random advection by a Kraichnan-Kazantsev velocity field. Combining tools from pattern formation with statistical theory and simulations, we show that random advection shifts the onset and the wave number of emergent patterns. As a simple model for pattern formation in convection, the effects are demonstrated with a generalized Swift-Hohenberg equation including random advection. We also discuss the implications of our results for the large-scale flow of turbulent Rayleigh-B\'enard convection.\\
\noindent \\
DOI: \href{https://journals.aps.org/prl/abstract/10.1103/PhysRevLett.123.114501}{10.1103/PhysRevLett.123.114501}
\end{abstract}

\maketitle
Many turbulent systems show a remarkable degree of large-scale coherence, despite the presence of strong fluctuations.
Large-scale convection patterns in the atmosphere and in the oceans are among the most fascinating examples. Rayleigh-B\'enard convection (RBC) \cite{siggia94arofm,bodenschatz00arofm,ahlers09rmp,lohse10arofm}, the flow between two plates heated from below and cooled from above, is a prototypical model for such flows and displays a range of phenomena\textemdash the emergence of laminar large-scale rolls close to the onset of convection, transitions to increasingly complex flow patterns as the temperature difference is increased, and finally, the emergence of turbulence. Close to onset, techniques like linear stability analysis as well as amplitude and phase equations explain the emergence and stability of convection patterns \citep{newell69jfm,segel69jfm,busse78ropp,pomeau80pla,cross84physica,cross93rmp,newell93arofm}.
Further away from the onset of convection, the flow becomes increasingly difficult to describe, especially when it becomes turbulent. Both experiments and, in particular, numerical simulations have provided insights into these complex flow regimes \cite{siggia94arofm,bodenschatz00arofm,ahlers09rmp}. Remarkably, coherent large-scale flow patterns, so-called turbulent superstructures, have been reported in the presence of small-scale turbulence \cite{hartlep03prl,parodi03prl,hartlep05jfm,vonhardenberg08pla,emran15jfm,stevens18prf,pandey18natcomm}. Using suitable averaging techniques, the topology and dynamics of the superstructures have been extracted from the turbulent flow fields, demonstrating, e.g.,large-scale dynamics reminiscent of spiral-defect chaos \cite{emran15jfm}. Extensive experimental and numerical investigations revealed that the length scale of the emerging patterns increases as a function of Rayleigh and Prandtl numbers as the flow becomes increasingly unsteady \cite{willis72jfm,heutmaker87pra,hu95pre} and, finally, turbulent \cite{hartlep05jfm,pandey18natcomm}. Investigations of RBC close to onset suggest that there is no universal scale selection mechanism, see, e.g., \cite{cross2009,getling98} for an overview. As soon as turbulence sets in, the issue is even more delicate: so far there is no conclusive explanation for the increased wavelength of turbulent superstructures. Interestingly, similar issues remain in explaining turbulent Taylor Couette flow \cite{grossmann16arofm}, i.e.~the flow between two rotating cylinders. Also here, the length scale of Taylor rolls increases with the Reynolds number \cite{ostilla16prf}, and the coexistence of large-scale flow states points at complex interactions between coherent flow and fluctuations \cite{huisman2014natcomm}.

On a more general level, this raises the fundamental question of how small-scale turbulent fluctuations impact the emergence and stability of large-scale flow patterns. Because of their complexity, a comprehensive explanation of these phenomena based on the full hydrodynamic equations appears formidable. However, as we demonstrate in this Letter, the role of turbulent fluctuations for pattern formation can be conceptually clarified by analyzing a much simpler, analytically tractable problem. We investigate a general class of pattern forming systems, in which the order parameter field is advected by a Kraichnan-Kazantsev velocity field \cite{kraichnan68pof,kazantsev68jetp}\textemdash a spatially correlated, white-in-time Gaussian field. Advection problems involving such fields have unveiled the role of small-scale fluctuations on the dynamo effect \cite{kazantsev68jetp}, and led to a better understanding of passive scalar turbulence (see \cite{falkovich01rmp} for a review). Coupling such a random advection field to a pattern-forming order parameter field allows us to analytically quantify its impact on the onset as well as the length scales of the emerging patterns. As a prototypical model for the emergence of large-scale patterns in convection, we illustrate our findings at the example of the Swift-Hohenberg (SH) equation \cite{swift77pra}, which we generalize to feature random advection. We find that random advection shifts the onset of pattern formation and effectively increases the pattern's wavelength by turbulent diffusion, offering a qualitative explanation for recent observations in turbulent RBC \cite{pandey18natcomm,stevens18prf}.

To start with, we consider a scalar order parameter field $\theta\ofx$ that exhibits pattern formation in two dimensions. Its nondimensionalized evolution equation takes the form
\begin{equation}
 \partial_t \theta + \bm{u} \cdot \nabla  \theta = \mathcal{L[\nabla ]}\theta + \mathcal{N}[\theta,\nabla ] \, .
  \label{eq:MechanismModel}
\end{equation}
Here, $ \mathcal{L}$ and $ \mathcal{N}$ denote linear and nonlinear operators, respectively. Additionally, the order parameter field is advected by a two-dimensional, zero-mean Gaussian velocity field $\bm{u}\ofx$ which is white in time and incompressible. As the Kraichnan-Kazantsev velocity field acts as a multiplicative advective noise, the resulting equation is a stochastic partial differential equation, which we interpret in the Stratonovich sense \cite{gardiner2010}; i.e., the velocity field is considered as rapidly varying in time in the limit of vanishing correlation time. The Gaussian random field has the correlation structure $\langle u_i\ofx u_j(\bm{x}',t')\rangle = {R_{ij}(\bm{x}-\bm{x}')\delta(t-t')}$, where the spatial covariance tensor takes the form $R_{ij}(\bm{\varrho}) = 2Q \big[ f(\varrho) \delta_{ij} + \varrho \left[\partial_\varrho f(\varrho)\right] \left\{\delta_{ij} - \hat{\varrho}_i \hat{\varrho}_j\right\} \big]$ as the result of homogeneity ($\bm \varrho = \bm x- \bm x'$), isotropy, and incompressibility. Here, $Q$ is the amplitude of the fluctuations, and $f$ denotes the longitudinal velocity correlation function with $f(0)=1$.

Within this setting, analytical statements can be made for the ensemble-averaged field.
To obtain such a description, we average Eq.~\eqref{eq:MechanismModel} with respect to realizations of the random advection. The mean advection term can be evaluated using the Furutsu-Donsker-Novikov identity (or Gaussian integration by parts) \cite{furutsu63jornbs,donsker64mit,novikov65jetp}:
{\fontsize{9.5}{11.4}\selectfont
\begin{equation}
\partial_i \langle u_i\ofx \theta\ofx\rangle = \partial_i \int\!d\bm x'\,R_{ij}(\bm x-\bm x')\left\langle\frac{\delta \theta(\bm x,t)}{\delta u_j(\bm x',t)}\right\rangle \, .
\end{equation}}
The mean response function is readily evaluated from the integral representation of Eq.~\eqref{eq:MechanismModel} resulting in $\left\langle \delta \theta(\bm x,t)/\delta u_j(\bm x',t)\right\rangle = -\frac{1}{2}\delta(\bm x-\bm x')\partial_j\langle \theta \rangle$. Therefore, we obtain
\begin{equation}
  \langle\bm{u}\ofx\cdot \nabla  \theta\ofx\rangle = -Q\Delta\langle\theta\rangle\ofx \, ,
  \label{eq:AveragedAdvection}
\end{equation}
where $\Delta=\nabla ^2$. On the level of the averaged field, only the random advection amplitude, and not the shape of the correlation function, plays a role.
As a result, the averaged equation takes the form
\begin{equation}\label{eq:meanfieldeq}
\partial_t \langle\theta\rangle = \left( \mathcal{L}[\nabla ] + Q \Delta \right) \langle\theta\rangle + \langle \mathcal{N}[\theta,\nabla ]\rangle.
\end{equation}
The important implication of this result is that the advection term contributes linearly to the averaged equation, adding an additional ``turbulent" diffusion. This means that eddy-diffusivity type closures, which have been proposed phenomenologically for RBC (see, e.g.,\cite{emran15jfm}), are exact for Kraichnan-Kazantsev velocity fields. Considering an ensemble-averaged field is only meaningful if a stationary large-scale pattern is achieved as its amplitude otherwise decays over time. As detailed in the Supplemental Material \cite{ibbeken19supplement}, this exact result based on ensemble averaging can be generalized to temporal coarse-graining, provided the coarse-graining scale is large compared to the scales of the random fluctuations but still smaller than the scales of the pattern formation. As a result, order parameter fields which vary slowly in time are also captured by this theoretical result.

To continue, we decompose the order parameter field into mean and fluctuations, $\theta = \langle \theta \rangle + \theta'$. If the random advection is comparably small in scale and amplitude, the assumption $\langle\mathcal{N}[\theta,\nabla ]\rangle \approx \mathcal{N}[\langle \theta \rangle,\nabla ]$ can be justified on the basis that the averages involving the fluctuations are small compared to the averaged field. This leads to a closed equation for the averaged field. The validity of this assumption is explicitly tested with simulations below, see Fig.~\ref{fig:examples}. If fluctuations around the averaged field need to be taken into account, they potentially yield additional linear and nonlinear contributions which can be treated separately.

\begin{figure}[!hbt]
 	\includegraphics[width=\columnwidth]{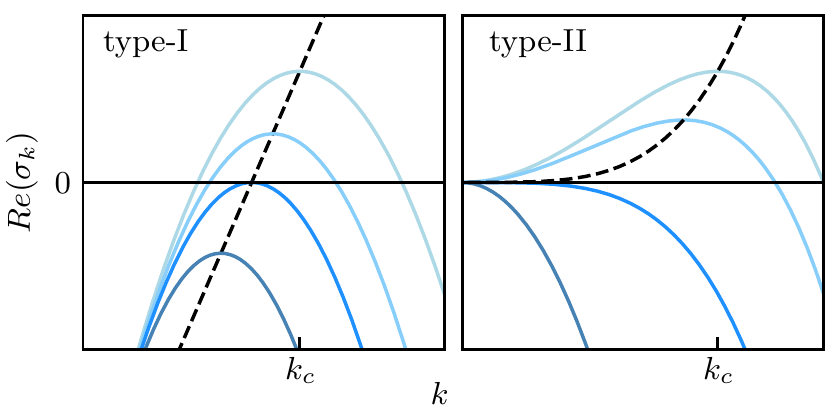}
 	\caption{\label{fig:bif_conceptual} Effect of random advection on type-I (left) and type-II (right) instabilities. As the random advection amplitude $Q$ is increased (from light to dark blue) for a fixed $r$, both the maximum growth rate and the corresponding wave number of the linear dispersion relation are shifted. Here, $k_c$ denotes the maximum-growth wave number of the case without advection ($Q=0$). The dashed line shows the maximum growth rate as a function of $k$ and $Q$.}
 \end{figure}

\begin{figure}[!hbt]
	\includegraphics[width=\columnwidth]{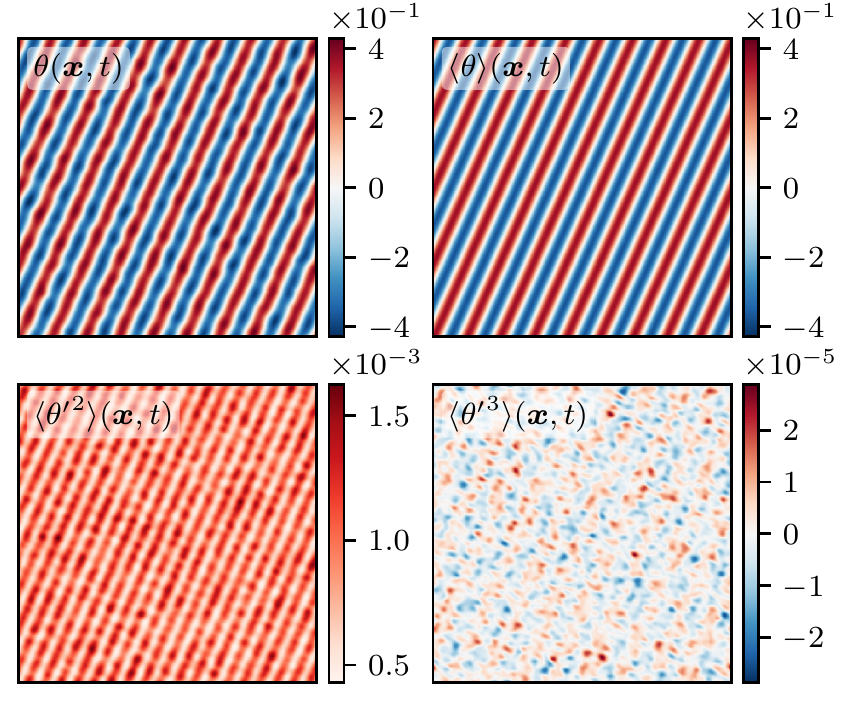}
	\caption{\label{fig:examples} Formation of a stripe pattern in the Swift-Hohenberg equation with random advection. Single realization (top left), ensemble-averaged field (top right), mean squared fluctuations (bottom left) and mean cubed fluctuations (bottom right) from simulations of Eq.~\eqref{eq:Model} in the statistically stationary state. Parameters are $\lambda=\lambda_c/10$, $Q=0.1$, and $r=0.2$. The simulations have been conducted on a periodic domain with $L=20\pi$ on a $256\times 256$ grid. A comparison of the field amplitudes shows that the fluctuations are negligible compared to the averaged field.}
\end{figure}

We proceed with a linear stability analysis by Fourier-expanding the averaged field. As is evident from Eq.~\eqref{eq:meanfieldeq}, the linear dispersion relation acquires an additional term $-Qk^2$, which shifts the position of the maximal linear growth rate. This can be made more transparent at the example of type-I and type-II instabilities \cite{cross93rmp,cross2009}. Both are characterized by a control parameter $\epsilon$ and a maximum-growth wave number $k_c$. For type-I instabilities, as in RBC, the linear dispersion relation is locally parabolic around the maximum-growth wave number. In the presence of random advection, the real part of the linear growth rate $\sigma_k$ takes the nondimensionalized form $\mathrm{Re} (\sigma_k) \approx \epsilon - (k-k_c)^2 - Qk^2$.
As a result, the critical control parameter is shifted from $\epsilon_c = 0$ to $ \epsilon_c=k_c^2Q/(1+Q)$,
and the critical wave number is shifted to a lower value $k_c^{\ast} = k_c/(1+Q)$.
For type-II instabilities, as they, for example, occur for conserved order parameters, the growth rate takes the form $\mathrm{Re} (\sigma_k) \approx  \epsilon k^2 - k^4/2  - Qk^2$, which shifts the critical control parameter from $\epsilon_c=0$ to $\epsilon_c=Q$ and implies a shifted maximum-growth wave number of $k_c^{\ast} = \sqrt{\epsilon - Q}$.
A sketch of the two types of instabilities is shown in Fig.~\ref{fig:bif_conceptual}, which illustrates the effect of the fluctuations.
As is evident from these considerations, random advection reduces 
the maximal linear growth rate and the corresponding wave number. For sufficiently strong
   advection, the onset of pattern formation can even be suppressed. 
The next step is a weakly nonlinear analysis in terms of amplitude equations. Provided the contributions of the fluctuations to the averaged equation are small, the structure of the amplitude equations is identical to the case without random advection, only the control parameter and the wave number need to be renormalized. Note that, depending on the specific model, the nonlinear pattern formation does not necessarily select the linearly predicted wave number \cite{cross2009}. In such cases, a more elaborate analysis is necessary.


To provide a concrete example relevant for convection and to further corroborate the analytical results, we investigate a generalized SH equation with random advection. The original SH equation has been introduced to model the onset of convection in the midplane of RBC in terms of an order parameter field, which can be interpreted as a linear combination of the temperature fluctuations and the vertical velocity \cite{swift77pra}. Although dramatically simplified compared to the Oberbeck-Boussinesq equations, it features the same type of instability and amplitude equations when only the slowest evolving modes of the full problem are considered \cite{cross80pof,ahlers81jfm}. By including random advection, we introduce a model for the fast, turbulent modes to the dynamics. The main motivation for this approach is the observation that the time-averaged patterns in the turbulent regime in RBC are reminiscent of the patterns close to onset, despite the presence of fluctuations \cite{emran15jfm,stevens18prf,pandey18natcomm}. Therefore the underlying assumption is that a description in terms of an order parameter field, superposed by fluctuations, remains valid in the turbulent case.
As we will see in the following, such a simple model suffices to conceptually explain recent observations on turbulent superstructures \cite{hartlep03prl,emran15jfm,stevens18prf,pandey18natcomm}. Suitably nondimensionalized, the SH equation with random advection takes the form
\begin{equation}
 \partial_t \theta + \bm{u} \cdot \nabla  \theta = [r-(\Delta+1)^2]\theta -\theta^3 \, .
  \label{eq:Model}
\end{equation}

Here, $r$ denotes the control parameter, which effectively models the Rayleigh number. In this nondimensionalization $k_c=1$, i.e.~$\lambda_c=2\pi/k_c=2\pi$. For the longitudinal velocity correlation we choose $f(\varrho) = \exp[-\pi \varrho^2/(4\lambda^2)]$ for simplicity. Therefore, the fluctuations are characterized by a single length scale $\lambda$. Alternative choices, e.g.,correlation functions corresponding to velocity fields with inertial-range scaling, are also possible. The spatial scales of the random fluctuations should be significantly smaller than the length scale of the large-scale pattern formation. Therefore, we choose $\lambda=\lambda_c/10$, which implies a scale separation, consistent with RBC in the soft-turbulence regime \cite{hartlep03prl}.

Assuming that the unclosed terms are small, we obtain the averaged equation
\begin{equation}
    \partial_t \langle \theta \rangle \approx \left [ r^{\ast} - \left( \Delta + k_c^{\ast 2}  \right)^2 \right ] \langle \theta   \rangle - \langle \theta   \rangle^3 \, .
	\label{eq:premean}
\end{equation}
In this approximation, the averaged equation takes exactly the form of the SH equation without advection, however, with renormalized control parameter and critical wave number. Consistent with the above discussion on the effects of random advection on type-I instabilities, we obtain a growth rate of $\sigma_k = r- (k^2-1)^2 - Qk^2$, which implies a shifted critical wave number $k_c^{\ast} := \sqrt{1-Q/2}$ and a shifted control parameter $r^{\ast} := r-Q+Q^2/4$. Corrections due to non-negligible fluctuations can be introduced by inclusion of the terms proportional to $\langle \theta'^2   \rangle$ and $\langle \theta'^3   \rangle$, which require additional closure. This can become necessary, for example, in the absence of scale separation between the random fluctuation and the large-scale pattern, leading to a breakdown of the approximation $\langle \mathcal{N}[\theta,\nabla ]\rangle \approx \mathcal{N}[\langle \theta \rangle,\nabla ]$. We characterize these limitations in the Supplemental Material \cite{ibbeken19supplement}.

The emergence of a pattern in the averaged field can be investigated by a single-mode amplitude equation which is readily derived from Eq.~\eqref{eq:premean} with the ansatz $\langle\theta\rangle\ofx= A(t) e^{\mathrm{i}\bm k_c^{\ast} \cdot \bm x} +\text{c.c.}$, where the direction of the renormalized wave vector $\bm k_c^{\ast}$ is arbitrary. The amplitude equation takes the form $\dot{A}(t) = r^{\ast} A(t) - 3 |A(t)|^2 A(t)$, which has the usual stationary solutions $|A(t)|=0$ for $r^{\ast} < 0$ and $|A(t)|=\sqrt{r^{\ast}/3}$ for $r^{\ast} \geq 0$, just with a renormalized control parameter. This implies that the bifurcation curves for different random advection amplitudes can be collapsed on one master curve, a prediction which we verify with simulations.

To this end, we rewrite Eq.~\eqref{eq:Model} in its It\^o 
formulation by applying the It\^o drift 
correction formula \cite{eyink02nl} and solve it by means of a pseudospectral method with a semi-implicit It\^o integrator for time stepping \cite{gardiner2010} with $\Delta t=0.1$.
Figure \ref{fig:examples} shows an example realization with $\lambda=\lambda_c/10$, $Q=0.1$, and $r=0.2$ along with the ensemble-averaged field and the mean squared and mean cubed fluctuations. The results for the ensemble-averaged field have been obtained by initializing 192 simulations of Eq.~\eqref{eq:Model} with identical initial conditions and letting them evolve with different realizations of the random advection field. Here, the initial conditions are stripe patterns with wave number $k_c^\ast$, which we let evolve under random advection to a statistically stationary state before we perform our analysis. While the realization and the averaged field are similar in amplitude, the unclosed terms involving fluctuations are orders of magnitude smaller, which gives an \textit{a posteriori} justification for neglecting them in Eq.~\eqref{eq:premean}.

\begin{figure}
	\includegraphics[width=\columnwidth]{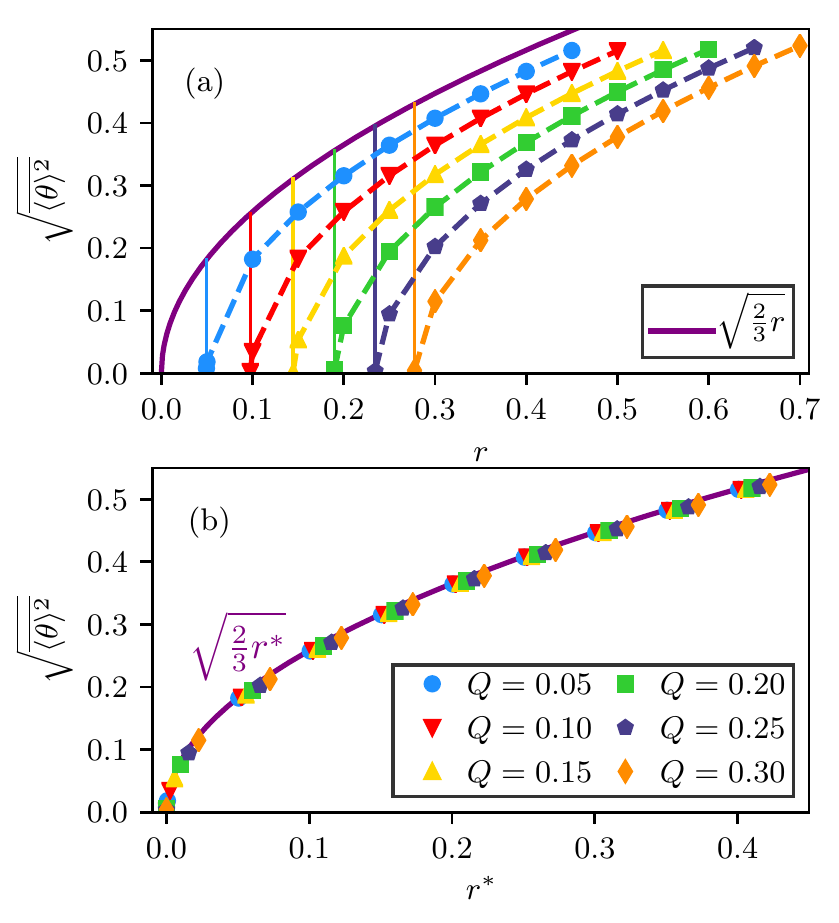}
	\caption{\label{fig:bifurcation}
		(a) Root mean squared amplitude in stationary state as a function of control parameter $r$ for varying amplitude $Q$ of the random advection field. All other simulation parameters are chosen as in Fig.~\ref{fig:examples}. Compared to the bifurcation curve without random advection (purple line), the onset is shifted. The theoretically predicted shift of onset (indicated with vertical lines) agrees well with the numerical results.
		(b) For the renormalized control parameter $r^\ast$, all cases collapse on the theoretically predicted master curve.}
\end{figure}

The theoretical predictions for the bifurcation in the presence of random advection are tested in Fig.~\ref{fig:bifurcation}. Plotted against the original, uncorrected order parameter, the onset of a large-scale pattern is shifted towards larger values of the control parameter. Renormalizing the control parameter confirms our theoretical result that the bifurcation scenario can be mapped onto the master curve $ \sqrt{\overline{\langle \theta \rangle^2}} = \sqrt{2r^{\ast}/3}$, where the overbar denotes spatial averaging. The factor of 2 compared to the amplitude equation result is a consequence of spatially averaging the order parameter field.

\begin{figure}
\includegraphics[width=\columnwidth]{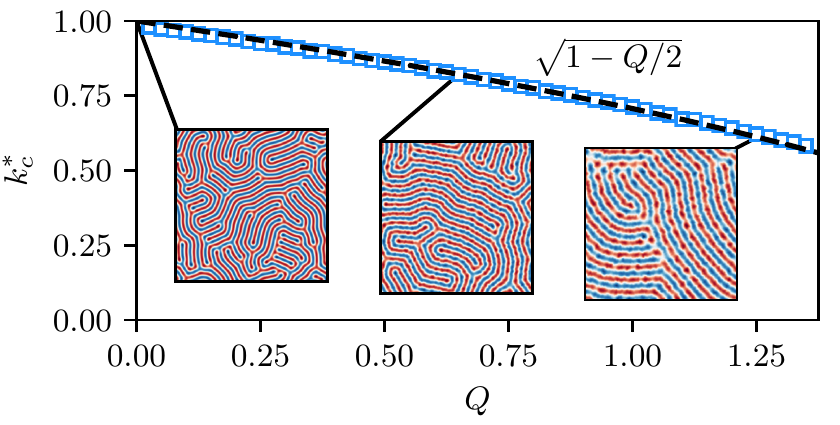}
\caption{\label{fig:wave number}Numerically obtained dominant wave number of the individual realizations (blue squares, simulations with $L=40\pi$ on a $512\times 512$ grid with $\lambda=\lambda_c/10$ and $r=0.9$) shown along with the theoretically predicted shifted critical wave number $k_c^{\ast}$ (dashed black line). As an illustration, the insets show individual realizations (scaled to unit maximum amplitude) for different $Q$.}
\end{figure}

So far, the theoretical discussion has focused on the averaged order parameter field. To test the predicted wave-number shift for a wider range of random advection amplitudes and to establish its validity beyond the ensemble with identical initial conditions, we investigate the randomly advected SH system significantly above onset ($r=0.9$) with random initial conditions. Figure \ref{fig:wave number} demonstrates excellent agreement of the theoretically predicted wave-number shift with simulation results. Analogous to the procedure established for RBC in \cite{pandey18natcomm}, the average wave number is obtained from the simulations from time-averaged spectra in the statistically stationary state, which corroborates that the characteristic shift is a feature of the individual realizations. This is also confirmed by the sample visualizations for various random advection amplitudes shown in the insets.

In conclusion, we have investigated the implications of random advection in the form of a Kraichnan-Kazantsev velocity field on a general class of pattern formation systems. We have shown that both the onset as well as the maximum-growth wave number are shifted in the presence of such fluctuations, which effectively introduce a turbulent diffusion. This effect is a direct consequence of the random advection and cannot be generated by standard additive or multiplicative noise.

Motivated by recent observations of turbulent superstructures in RBC, we investigated a generalized SH equation with random advection to mimic the impact of small-scale turbulent fluctuations on the emergence of large-scale patterns. Sufficiently close to onset, the bifurcations for various random advection amplitudes can be collapsed onto one master curve if the control parameter and the critical wave number are renormalized. We also confirmed our theoretical prediction of a renormalized critical wave number significantly above onset.

Let us briefly set our findings into the context of previous work on stochastic SH models.
Additive noise, which has previously been considered \cite{swift77pra,hohenberg92pra,elder92prl}, for example, to account for the impact of thermal fluctuations, does not yield any contribution on the level of the averaged equation. Multiplicative noise in the form of a fluctuating control parameter leads to a variety of effects \cite{garcia12}: For example, \citet{ojalvo93prl} observed a noise-induced onset of pattern formation, i.e.~an effective shift of the critical control parameter to smaller values, for noise which rapidly fluctuates in space and time. For spatially constant multiplicative noise, on-off intermittency \cite{platt93prl,aumaitre05prl} has been observed in SH models \cite{fujisaka01pre}. Random advection to model turbulent fluctuations, as considered here, adds a fundamentally different aspect: The critical wave number of the arising pattern is shifted to smaller values, and the critical control parameter is shifted to larger ones.

In RBC, the Reynolds number is a function of the Rayleigh number and the Prandtl number \cite{siggia94arofm,ahlers09rmp}. In contrast to that, the control parameters $r$ (modeling the Rayleigh number) and $Q$ (amplitude of turbulent fluctuations) can be varied independently in our model. This allowed the detailed investigation of the role of fluctuations on the emergence of patterns, which cannot be directly studied in RBC. With turbulent diffusion leading to a larger wavelength of the emerging pattern, our model offers a qualitative explanation for the wavelength growth of the turbulent superstructures with an increasing Rayleigh number \cite{hartlep05jfm,pandey18natcomm,stevens18prf}. Therefore, the presented results may guide future quantitative investigations of the role of turbulent fluctuations in RBC and other turbulent flows with emergent large-scale patterns.

\begin{acknowledgments}
This work is supported by the Priority Programme SPP 1881 Turbulent Superstructures of the Deutsche Forschungsgemeinschaft. We thank Wouter J.~T.~Bos, Ragnar Fleischmann, Martin James and Stephan Weiss for helpful discussions and the anonymous referees for their comments, which helped to improve the manuscript.

G.I. and G.G. contributed equally to this work.
\end{acknowledgments}

\bibliography{random_pattern_formation}

\renewcommand{\thefigure}{S\arabic{figure}}
\renewcommand{\theequation}{S\arabic{equation}}
\stepcounter{myequation}
\stepcounter{myfigure}
\onecolumngrid
\pagebreak
\newpage 
\begin{center}
	\textbf{\large Supplemental Material for `Large-Scale Pattern Formation in the Presence of Small-Scale Random Advection'}\\[.2cm]
	Gregor Ibbeken,$^{1,2}$ Gerrit Green,$^{1,2}$ and Michael Wilczek$^{1,2}$\\[.1cm]
	{\itshape ${}^1$	Max Planck Institute for Dynamics and Self-Organization (MPI DS),\\
		Am Fa\ss berg 17, 37077 G\"ottingen, Germany\\
		${}^2$Faculty of Physics, University of G\"{o}ttingen, Friedrich-Hund-Platz 1, 37077 G\"{o}ttingen, Germany}\\[1cm]
\end{center}
\twocolumngrid

In this Supplemental Material, we discuss a generalization of the ensemble averaging approach presented in the main text to running time averages, which allows the application to slowly evolving patterns. Furthermore, we provide more insights into the limitations of our approach.

\section{Generalization to time averaging}
In the following, we show numerically that the same results as for the ensemble average
can be obtained using a running time average. This allows us to extend our analysis to large-scale patterns that vary slowly in time. This is an important generalization since turbulent superstructures, e.g., in Rayleigh-B\'enard convection, also slowly evolve with time \cite{emran15jfm,pandey18natcomm}. Here, we limit ourselves to the Swift-Hohenberg equation with random advection since the generalization to other pattern-forming systems is straightforward.

To start with, we introduce the time-averaged order parameter field
\begin{equation}
\langle{\theta}\rangle_T(\bs x ,t) = \frac{1}{T}\int_{t-T/2}^{t+T/2} \!\!\!\! \theta(\bs x,s) \, ds \label{eq:timeavg}.
\end{equation}
Here, $T$ is the time scale over which the running average is taken. In order to remove the fluctuations, $T$ has to be chosen significantly larger than their characteristic time scale.
Time-averaging Eq.~(5) from the main text leads to 
\begin{equation}
\partial_t\langle{\theta}\rangle_T+\nabla\cdot\langle\boldsymbol{u}\theta\rangle_T=[r-(\Delta+1)^2]\langle{\theta}\rangle_T-\langle\theta^3\rangle_T \label{eq:mean_T},
\end{equation}
where we have used the incompressibility condition ${\nabla\cdot\bs u=0}$.
\begin{figure}
	\centering
	\includegraphics[width=1.0\linewidth]{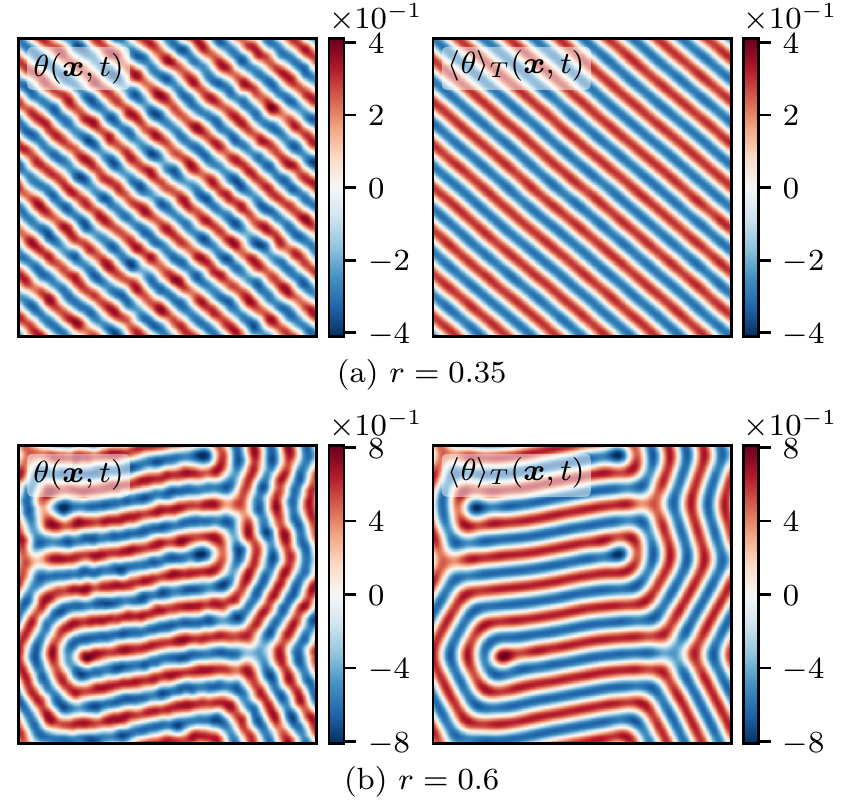}
	\caption{\label{fig:bif_time_real}Formation of a large-scale pattern in the Swift-Hohenberg equation with random advection. Single realization (left), time-averaged field (right). Parameters are $\lambda=\lambda_c/10$, $Q=0.3$, (a) $r=0.35$ and (b) $r=0.6$. The simulations are conducted on a periodic domain starting from random initial conditions. Because of the random initial condition, a complex large-scale pattern emerges in (b).}
\end{figure}
Because the random velocity field evolves rapidly in time (it is delta-correlated) while the large scales evolve slowly, we assume that the time average can be replaced by an ensemble average.
This allows to apply the Furutsu-Donsker-Novikov identity \cite{furutsu63jornbs,donsker64mit,novikov65jetp}, which, of course, leads to precisely the same evolution equation as for the case of the ensemble average. 

\begin{figure}
	\centering
			\includegraphics[width=1.0\linewidth]{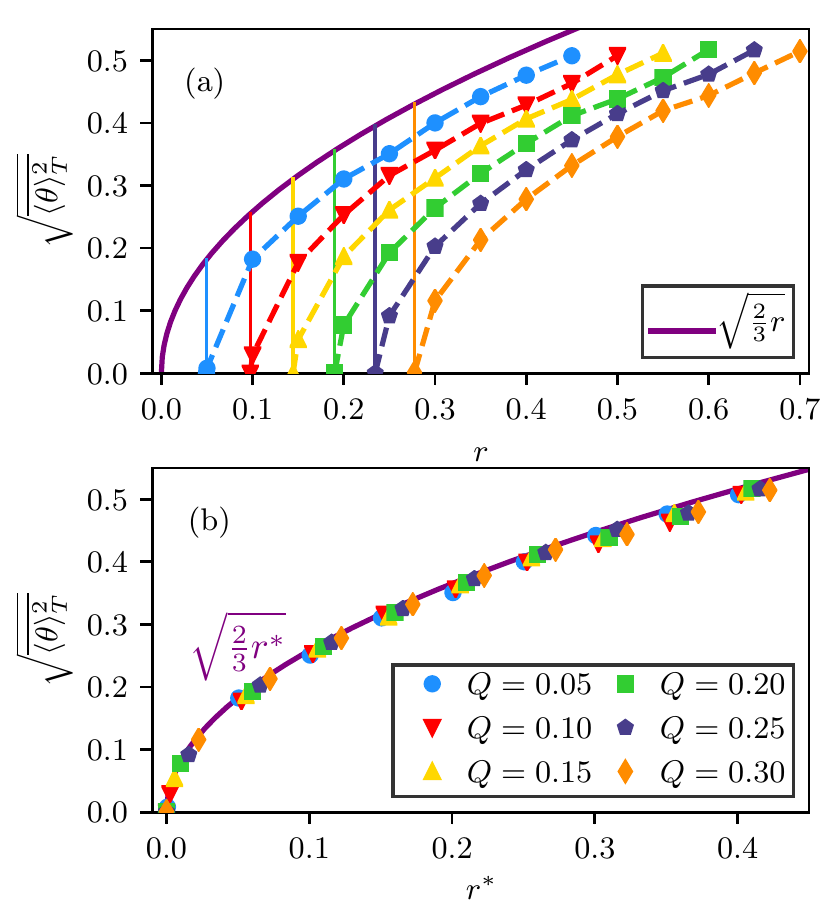}
			\caption{\label{fig:bif_time}
					(a) Root mean squared amplitude of the time-averaged order parameter field in the stationary state as a function of control parameter $r$ for varying amplitude $Q$ of the random advection field.
					(b) For the renormalized control parameter $r^\ast$, all cases collapse on a master curve, matching the theoretically predicted curve very well.}
\end{figure}

Similar to the procedure discussed in the main text, the order parameter field can be decomposed into its mean and fluctuations, $\theta = \langle{\theta}\rangle_T + \theta'$. For the cases considered, one can confirm from the simulations that the unclosed fluctuations in Eq.~\eqref{eq:mean_T}, originating from the cubic nonlinearity in the Swift-Hohenberg equation, are small compared to the time-averaged field (see also discussion in the following section).
This justifies $\langle\theta^3\rangle_T\approx \langle\theta\rangle_T^3$, and, therefore, we obtain
\begin{equation}
\partial_t\langle{\theta}\rangle_T\approx[r^{\ast}-(\Delta+k_c^{\ast 2})^2]\langle{\theta}\rangle_T -\langle{\theta}\rangle_T^3,
\end{equation}
where $r^{\ast}=r-Q+Q^2/4$. As a result, the following analysis is analogous to the one presented in the main text. Accordingly, we find that the onset is shifted from $r_c=0$ to $r_c=Q-Q^2/4$ and that the critical wave number is reduced to $k_c^\ast=\sqrt{1-Q/2}$. For the amplitude, we also obtain $\sqrt{\overline{\langle \theta\rangle_T^2}} = \sqrt{2 r^\ast/3}$, as expected. To verify this result, we perform simulations like the ones presented in the main text with $N_x=N_y=256, L=20\pi$, and $\Delta t=0.1$. The averaging window is $T=500$, and the time difference is $\Delta s=1$ between snapshots for the evaluation of Eq.~\eqref{eq:timeavg}. In contrast to the simulations in the main text, these simulations are started from random initial conditions instead of an ideal stripe pattern. It is important to note that the temporal averaging window has to be chosen carefully. It has to be long enough in order to remove the small-scale fluctuations but significantly shorter than the time scale 
on which the large-scale pattern evolves.
This is similar to the procedure applied to Rayleigh-B\'e{nard} convection \cite{emran15jfm,pandey18natcomm}. Example snapshots and averaged fields for different $r$ are shown in Fig.~\ref{fig:bif_time_real}. Even when started from random initial conditions, a non-vanishing large-scale pattern emerges in the time-averaged field. We then obtain the full bifurcation diagram from time-averaged simulations and confirm that it matches the theoretical prediction, see Fig.~\ref{fig:bif_time}. Compared to the theoretical expectation, only minor deviations can be seen due to the random initial conditions. Therefore, we conclude that our findings can be generalized to running time averages of fields with slowly evolving large scales.

\section{Limitations of the approach}
\begin{figure}
	\centering
	\includegraphics[width=1.0\linewidth]{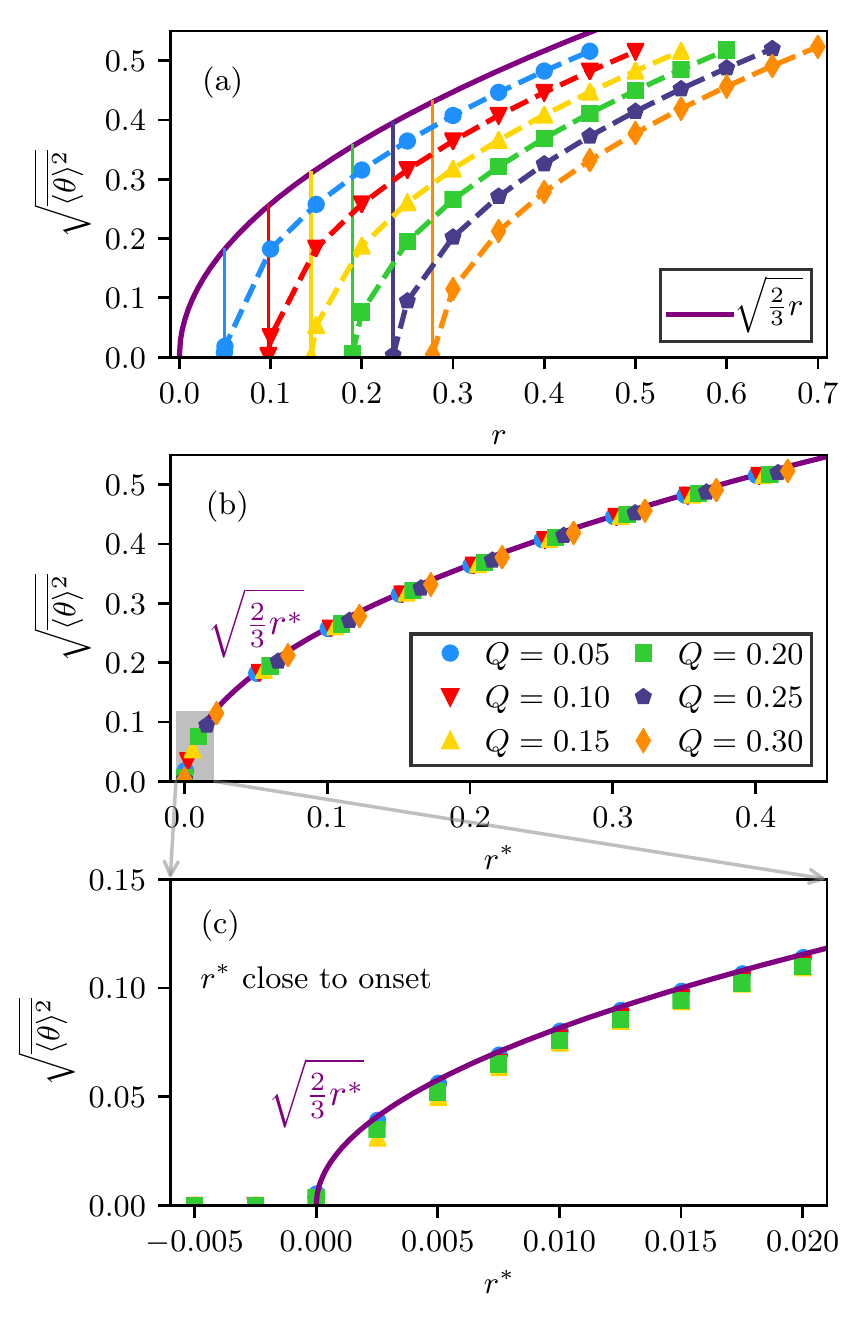}
	\caption{\label{fig:bif_onset} Root mean squared amplitude of the ensemble-averaged field as a function of (a) $r$ and (b) $r^\ast$. Panels (a) and (b) are identical to Fig.~3 in the main text. Panel (c) shows a zoom into the range close to onset with additional data from longer simulations, cf.~Fig.~\ref{fig:bif_time_onset}. Simulation details are given in the main text.}
\end{figure}

In this section, we discuss the limitations of our approach. 
To this end, consider the approximation made in the main text:
\begin{equation}\label{eq:approximation}
  \langle\mathcal{N}[\theta,\nabla]\rangle \approx \mathcal{N}[\langle \theta \rangle,\nabla].
\end{equation}
Here, $\langle \cdot \rangle$ denotes an ensemble average. When we introduce the decomposition $\theta = \langle \theta \rangle+\theta^\prime$ and average the Swift-Hohenberg equation with random advection, we obtain
\begin{equation}
  \partial_t \langle \theta \rangle = \left [ r^{\ast} - \left( \Delta + k_c^{\ast 2}  \right)^2 \right ] \langle \theta   \rangle -\langle\theta^3\rangle,
\end{equation}
where
\begin{equation}
  \langle\theta^3\rangle = \langle \theta   \rangle^3 + 3 \langle \theta'^2 \rangle\  \langle \theta \rangle +3\langle \theta' \rangle \langle \theta \rangle^2 +\langle \theta'^3 \rangle \, .
\end{equation}
\begin{figure}[!hbt]
	\centering
	\includegraphics[width=1\linewidth]{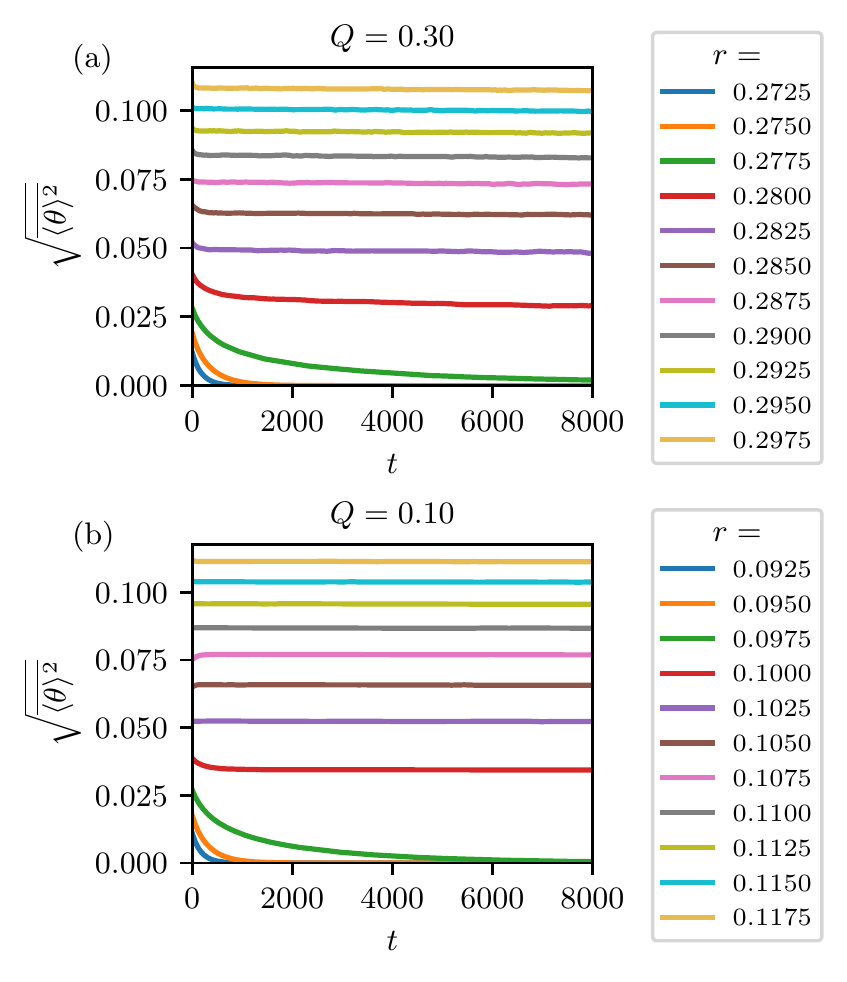}
	\caption{\label{fig:bif_time_onset} Time dependence of the root mean squared amplitude of the ensemble-averaged field for different $r$ with (a) $Q=0.3$ and (b) $Q=0.1$ very close to onset from long-time simulations. (a) For large $Q=0.3$, the amplitude of the unclosed fluctuations is not small compared to the order parameter field. Here, an extremely slow decay of the amplitude of the averaged field can be observed. The three curves with the fastest decay correspond to the cases for which $r^{\ast}\leq0$, in agreement with our theoretical prediction. (b) For small $Q=0.1$, the mean amplitude does not decay over time for all parameters above onset. Only for the cases with $r^{\ast}\leq0$, the amplitudes decay, as expected.}
\end{figure}
The term $3\langle \theta' \rangle \langle \theta \rangle^2$ vanishes because $\langle \theta' \rangle = 0$, but $3 \langle \theta'^2 \rangle \langle \theta \rangle$ and $\langle \theta'^3 \rangle$ remain. However, as illustrated in Fig.~2 of the main text, these terms are very small in all cases of relevance, which justifies our approximation \eqref{eq:approximation}. This is also reflected in the remarkable overall agreement between the theoretical prediction and the numerical results in Fig.~3 of the main text as well as in Fig.~\ref{fig:bif_onset}. Compared to the main text, Fig.~\ref{fig:bif_onset}(c) additionally shows a zoom into the region very close to onset whereas panels (a) and (b) are identical to Fig.~3.

\paragraph{Deviations close to onset}
\begin{figure}[!hbt]
	\centering
	\includegraphics[width=1.0\linewidth]{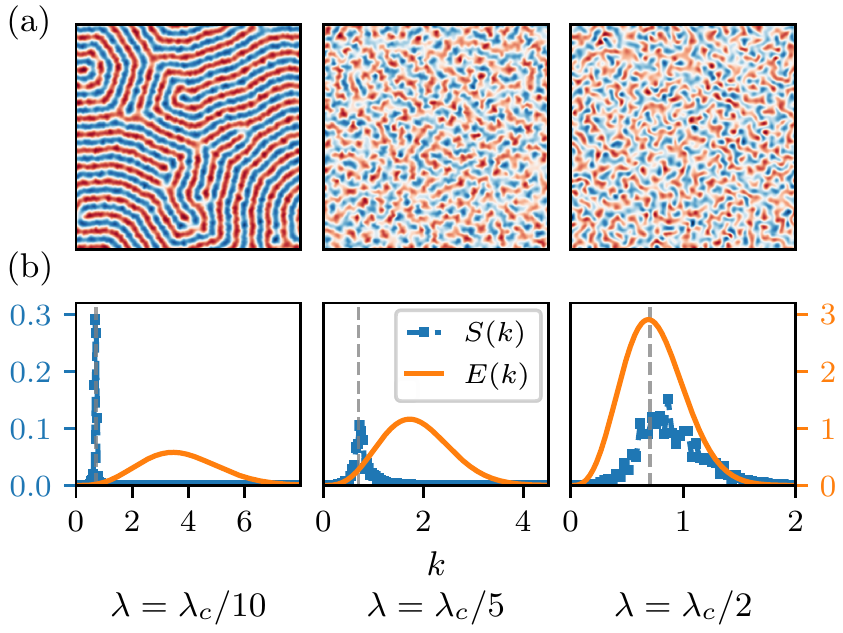}
	\caption{(a) Realization of $\theta$ (scaled to unit maximum amplitude) for $r=0.9,~Q=1.0$, and different $\lambda$. With increasing length scale, the fluctuations suppress the emergence of a large-scale pattern. (b) Corresponding azimuthally averaged spectrum $S(k)$ of $\theta$ compared to the energy spectrum $E(k)$ of the Gaussian random velocity field. A clear scale separation between the spectra is visible for $\lambda=\lambda_c/10$, which vanishes with increasing $\lambda$. The theoretical expectation for the location of the peak, $k_c^\ast$, is indicated by the dashed line. For the largest length scale $\lambda$, we observe a deviation of the peak from the theoretical expectation. Simulations are performed on a $512\times512$ grid with $L=40\pi$ and started from random initial conditions.}
	\label{fig:SH_l}
\end{figure}

The root mean squared amplitude of the averaged field is $\sqrt{2 r^\ast/3}$ and, therefore, is small itself close to onset. This may very well lead to the breakdown of the approximation \eqref{eq:approximation} as the previously neglected fluctuation terms gain importance. 
To check the validity of the approximation close to the bifurcation point ($r_c^\ast=0$), we map out the bifurcation diagram in this regime, which is presented in Fig.~\ref{fig:bif_onset}(c). 
Minor systematic deviations can be observed. For small random advection amplitudes, these can be accounted for by including the neglected fluctuations.
However, given the still excellent agreement, we choose to apply the approximation \eqref{eq:approximation} which leads to a closed equation for the averaged field and allows to derive the renormalized control parameter and wave number in a predictive manner.

More generally, the breakdown of our ensemble averaging approach is reflected by the fact that a stationary non-vanishing averaged field does not develop. As illustrated by Fig.~\ref{fig:bif_time_onset}(a), this is the case very close to onset for $Q=0.3$. Here, a
slow decay of the averaged field's amplitude can be observed. However, this is only the case for the largest amplitudes of the random advection field and very close to onset, where the amplitude of the order parameter field is small. For smaller $Q$, the amplitude is stationary even close to onset, see Fig.~\ref{fig:bif_time_onset}(b) for $Q=0.1$. For the cases considered in our main text, the amplitude is stationary even close to onset. In Fig.~\ref{fig:bif_onset}(c), we only show data for random advection amplitudes $Q$, for which the root mean squared amplitudes of the ensemble-averaged fields are stationary for all $r$. As a result, we exclude the cases $Q=0.3$ and $Q=0.25$, for which the points corresponding to the two values of $r^\ast$ directly above onset are not stationary.

\paragraph{Large velocity correlation length}
Finally, we briefly discuss the influence of the length scale $\lambda$ (see main text for details) characterizing the correlation length of the velocity fluctuations. The approximation \eqref{eq:approximation} requires scale separation between fluctuations and the large-scale pattern. If this scale separation is absent, the unclosed terms in the averaged equation play a crucial role and cannot be neglected \textit{a priori}. In extreme cases, stationary or slowly evolving large-scale patterns do not even develop. Different realizations then vary strongly and, therefore, cancel on average, prohibiting an application of our theoretical results. Indications for this can be seen in Fig.~\ref{fig:SH_l}, in which a comparison for different length scales $\lambda$ is shown. Here, the large-scale pattern smears out at large $\lambda$. In order to quantify the scale separation, we compare the azimuthally averaged spectrum $S(k)$ of $\theta$ with $E(k)$, the prescribed spectrum of the Gaussian random velocity field, in Fig.~\ref{fig:SH_l}. The spectra $S(k)$ show a single peak for all $\lambda$, but their width grows with $\lambda$. In contrast, the velocity spectrum $E(k)$ becomes narrower. For the smallest considered correlation length $\lambda$, the peak of the spectrum of $\theta$ is at $k_c^\ast$, separated from the peak of $E(k)$ at much larger wave numbers. Here, a stationary large-scale pattern emerges. This illustrates how scale separation between the large-scale pattern and the most intense small-scale fluctuations is essential.

To summarize, in this Supplemental Material we have shown that our results are not limited to the case of ensemble averaging, but also hold for running time averages. Additionally, we have provided insights into the limitations of our approach and showed that it only starts to fail very close to onset and for a large correlation length of the random velocity field.

\end{document}